\documentclass[doublecol]{epl2_nodraft}
\usepackage[english]{babel}
\usepackage{epsfig,latexsym,amsmath,mathrsfs,amssymb}
\usepackage{xcolor}
\usepackage[colorlinks]{hyperref}
\usepackage[nameinlink]{cleveref}

\def \nn{\nonumber}
\def \be{\begin{equation}}
\def \ee{\end{equation}}

\title{Hawking radiation from acoustic black holes in hydrodynamic flow of electrons}

\author{
Shreyansh S. Dave\inst{1}\thanks{E-mail: \email{shreyansh.dave@tifr.res.in}} 
\and Oindrila Ganguly\inst{2}\thanks{Email: \email{oindrilacg@gmail.com}} 
\and Saumia P.S.\inst{3}\thanks{Email: \email{saumia@gmail.com}} 
\and Ajit M. Srivastava\inst{4,5}\thanks{Email: \email{ajit@iopb.res.in}}
}

\shortauthor{S. S. Dave \etal}

\institute{
\inst{1} Department of Nuclear and Atomic Physics, Tata Institute of Fundamental Research, Mumbai 400005,  India \\
\inst{2} Department of Physics, Indian Institute of Science, Bengaluru 560012, India \\
\inst{3} Bogoliubov Laboratory of Theoretical Physics, JINR, 141980 Dubna, Russia \\
\inst{4} Institute of Physics, Bhubaneswar 751005, India, \\
\inst{5} Homi Bhabha National Institute, Training School Complex, Anushakti Nagar, Mumbai 400085, India
}

\abstract{
Acoustic black holes are formed when a fluid flowing with subsonic velocities, accelerates and becomes supersonic. When the flow is directed from the subsonic to supersonic region, the surface on which the normal component of fluid velocity equals the local speed of sound acts as an acoustic horizon. This is because no acoustic perturbation from the supersonic region can cross it to reach the subsonic part of the fluid. One can show that if the fluid velocity is locally irrotational, the field equations for acoustic perturbations of the velocity potential 
are identical to that of a massless scalar 
field propagating in a black hole background. One, therefore, expects 
Hawking radiation in the form of a thermal spectrum of phonons. There have been numerous investigations
of this possibility, theoretically, as well as experimentally, in systems
ranging from cold atom systems to quark-gluon plasma formed in relativistic
heavy-ion collisions. Here we investigate this possibility in the hydrodynamic
flow of electrons. Resulting Hawking radiation in
this case should be observable in terms of current fluctuations. Further, 
current fluctuations on both sides of the acoustic horizon should show 
correlations expected for pairs of Hawking particles.
}

\begin{document}

\maketitle

Laboratory analogues of cosmic/astrophysical phenomena have proved to be
of great importance. One of the most important examples of this is the
laboratory analogue of black holes, the so called acoustic 
black hole \cite{unruh1981,visser1998}.
Black holes are probably the most exotic objects known to occur in the
Universe. There is ample observational evidence of accretion disks around
astrophysical black holes, but the region close to the horizon has 
not been readily accessible. (Though,
with gravitational wave observations of black hole collisions, even this
regime of black hole physics should be within reach of 
future experimental investigations). Laboratory investigations with
acoustic black holes can be very useful to get insight into this regime.
Probably the most intriguing phenomenon associated with black holes is
Hawking radiation arising from the behavior of quantum fields in
the background of a black hole spacetime \cite{hawking1975,robertson2012}. 
It does not seem possible, in
any foreseeable future experiment, to probe this phenomenon for any
astrophysical  black hole which typically has a Hawking temperature less than
about 10$^{-7}$ K (for a stellar mass black hole), much smaller than the 
temperature of cosmic microwave
background radiation. It is well appreciated that Hawking  radiation
raises deep conceptual issues related to unitary evolution and 
information loss in the 
formation and subsequent evaporation of black holes. Any experimental
probe of the physics of Hawking radiation will be an important step
towards understanding this important phenomenon. It is not surprising
that many investigations with acoustic black holes have focused
on the possibility of observing Hawking radiation in these laboratory
analogues.

An acoustic black hole is a specific case of a more general result
related to the propagation of acoustic perturbations in the velocity 
potential of an inviscid, barotropic fluid. It was shown by Unruh that such
acoustic perturbations obey an equation which is identical to the Klein 
Gordon equation for a massless scalar field in a curved Lorentzian spacetime 
\cite{unruh1981}, with the spacetime metric determined by the
flow  velocity, density, and pressure of the fluid. An $n$ dimensional analogue system gives rise to an $(n+1)$ dimensional analogue spacetime.
As the acoustic
perturbations propagate with speed of sound in the fluid, it is clear
that if on a surface, the normal component of fluid velocity equals
the local speed of
sound, and becomes supersonic beyond it, then
no acoustic perturbation  can cross this surface from the supersonic
region to the subsonic region. One can then expect such a flow
geometry to correspond to a black hole with this special
surface being identified with the horizon of the black hole.
Indeed, as shown by Unruh \cite{unruh1981}, for  a spherically symmetric,
stationary, convergent background fluid flow, one finds the effective
metric seen by acoustic perturbations of the velocity potential
to be the Schwarzschild metric, with the horizon
coinciding with the surface where the fluid velocity becomes supersonic. 
It was then predicted in \cite{unruh1981} that in a fluid where acoustic perturbations
can be quantized, one should expect Hawking
radiation in terms of thermal bath of acoustic phonons emitted from
this sonic horizon.

Numerous studies have been carried out to probe this 
possibility (\cite{barcelo2005,novello2002book,garay1999,lahav2009,drori2018,guo2019,bera2020,blencowe2020} and references therein).                     
Many investigations with cold atom systems have focused on the signature 
of Hawking radiation in terms of correlated pairs of 
Hawking particles emitted from the sonic horizon, with the two partners of
the pair propagating on the two sides of the sonic horizon
\cite{carusotto2008,macher2009,steinhauer2015,steinhauer2016,denova2018}. Interestingly,
though one can calculate properties of correlations among such pairs for
the Hawking radiation of a real black hole, its experimental investigation
is simply out of the question as the region inside the event horizon is causally
disconnected from the physically accessible region outside the horizon. 
For acoustic black holes, in contrast, it is simply a matter of observing
acoustic perturbations on the two sides of the sonic horizon, with both
sides equally accessible to experiments. It has been claimed that the 
observations are in agreement with the theoretical predictions for
Hawking radiation. Such observations are very important, providing first 
ever experimental evidence of the basic physics underlying Hawking radiation.
It will be highly desirable
to find some experimental situation where the Hawking radiation
can be observed directly in terms of thermal spectrum of acoustic phonons. There have
been some investigations in this direction \cite{barcelo2005}.                                                
It has also been proposed by some of us that an acoustic black hole metric
may be constructed in the flow of quark-gluon plasma (QGP) in relativistic
heavy-ion collisions \cite{das2020}. In that case, the resulting thermal 
radiation of acoustic phonons may be observable in terms of modification of
 the rapidity dependence of the transverse momentum distribution of 
various particles.

In this work, we propose another possible analogue model for acoustic black holes
where resulting Hawking radiation may be observable directly as thermal
radiation of emitted phonons. We consider hydrodynamic flow
of electrons. Possibility of electron hydrodynamics was first proposed
by Gurzhi \cite{gurzhi1963,gurzhi1968} for a system where electron-electron scattering
dominates over momentum non-conserving scattering 
of electrons, e.g. with impurities
and with phonons. Electron-electron scattering conserves the net momentum of
electron system thus leading to conservation equations, namely the
hydrodynamical equations for electron flow. Theoretically, it is a clean
argument, but the situation with experiments has not been so clean. It took
several decades to achieve ultra-clean systems where this regime of
dominant electron-electron scattering could be achieved. Hydrodynamical
flow of electrons is believed to have been achieved in a quasi 2-dimensional
electron gas  in high mobility heterostructures (e.g. (Al,Ga)As 
heterostructures \cite{molenkamp1994,dejong1995}) , in graphene \cite{bandurin2018,lucas2018,narozhny2017},
as well as in Dirac and Weyl semimetals in 3-dimensions \cite{gooth2018}.  
In such systems, observations related to viscous effects of the Navier-Stokes' equation in electronic transport, such as 
Poiseuille-like flow profile, flow pattern of vortices, etc. have been reported. There are also
proposals for probing non-linear hydrodynamical effects e.g. Bernoulli effect, 
Eckart streaming, and Rayleigh streaming  of vortices \cite{hui2020}. 

We will focus on an entirely different aspect of hydrodynamical flow of 
electrons. We will consider specific geometry of the sample which allows 
the flow to become supersonic beyond a surface. To be specific, we will
consider example of a quasi $2$-dimensional electron gas, e.g. in ultra-clean
heterostructures, assuming the system to have sufficient
thickness that it may be treated as $3$ dimensional. This allows us to establish
correspondence with a $(3+1)$ dimensional black hole. We will then write down the 
analogue black hole metric and estimate resulting Hawking temperature for
specific system parameters. We will argue that the Hawking temperature in
this system will manifest in terms of electric current oscillations with
thermal spectrum which may be observable. We mention that for Hawking 
radiation from an acoustic black hole, it is important that the fluid should
have quantum nature as Hawking radiation results from the quantized modes
of the relevant field. This is what is achieved in Bose Einstein condensate (BEC) systems \cite{garay1999,carusotto2008,macher2009} and in the                         
proposed quark gluon plasma (QGP) system produced in relativistic heavy-ion collisions \cite{das2020}. 
This is also true for the present electron-hydrodynamics system expected to be manifest in ultra-clean systems with
strong quantum correlations. Please note that our results can be smoothly extended to lower dimensions. We could equally well have taken the sample to be exactly 2 dimensional, e.g. graphene, in which case the resulting analogue black hole spacetime would have been $(2+1)$ dimensional. The occurrence of Hawking radiation and the estimate of Hawking temperature rely on the behaviour of quantum fields in a curved spacetime with appropriate boundary conditions and so remain valid in lower dimensions too.

We  begin by briefly reviewing how an acoustic metric can be constructed in a non-relativistic fluid and argue that this smoothly applies to the case of electron transport in the hydrodynamic regime. Next, we write down the expression for Hawking temperature of an acoustic black hole formed in a suitably chosen model of one dimensional fluid flow. Following this, we discuss sample geometries, in particular that of a de Laval nozzle, that can give rise to flow configurations necessary for formation of acoustic black holes and rewrite Hawking temperature in terms of shape of the
de Laval geometry. Taking typical parameter values available in literature, we give an estimate of the Hawking temperature and the amplitude and frequency spectrum of resulting current oscillations. We also discuss
the possibility of observing Hawking pair correlations in such a system
in terms of current-current correlations between two sides of the acoustic
horizon. Finally, we conclude with a discussion of the limitations of our analysis and various future possibilities.


\section{Hydrodynamics of electrons and acoustic black hole metric} \label{sec:rev}
Possibility of a hydrodynamic regime for electron transport was first 
 discussed by Gurzhi \cite{gurzhi1963,gurzhi1968} for ultra-clean crystals
where electron-electron scattering, which conserves the momentum of 
the electron fluid, dominates over the scattering of electrons
 with impurities and with phonons which do not preserve
the momentum of electrons. As mentioned in the Introduction, this 
remarkable possibility has been realized in
some ultra-clean systems (to suppress electron-impurity scattering) 
at appropriately low temperatures (to suppress electron-phonon scatterings, 
but still allowing significant electron-electron scattering). There have
been numerous investigations 
\cite{gurzhi1963,gurzhi1968,molenkamp1994,dejong1995,lucas2018,narozhny2017,bandurin2018,gooth2018,hui2020,moors2021}
discussing constraints on impurity  
concentration and the regime of temperature that make
electron-electron scattering dominant over the other, momentum 
non-conserving, scattering modes. One also requires constraints on the system 
size as scattering of electrons with the boundaries of the sample, in 
general, leads to momentum loss from the electron fluid. With sufficient evidence 
available for validity of this regime in these ultra-clean materials,
we assume validity of hydrodynamic description of electron 
transport in these systems and write down the Navier-Stokes' equations
for the electron fluid. An important point we 
mention here is that we restrict our discussion to the case of an inviscid
fluid. While experiments show that electron hydrodynamical flow has
non-zero viscosity, the viscous  effects do not appear to be dominant compared
to other relevant effects like scattering of 
electrons with impurities, phonons, and the sample boundary.
For certain cases, e.g. graphene, viscous effects may actually be 
negligible \cite{moors2021}.  

For non-relativistic hydrodynamics the basic equations of fluid flow are the
following: the continuity equation
\begin{equation}
\partial_t \rho + \vec\nabla . (\rho {\vec v}) = 0
\end{equation}
and the Euler equation
\begin{equation}
\rho [\partial_t {\vec v} + ({\vec v}.\vec\nabla) {\vec v}] + \vec\nabla p = 0 ~.
\end{equation}
This is in the absence of any external force. For electron fluid, there
should be a term involving electrostatic potential. We  are not including 
that for simplicity. Such a term can be absorbed in the pressure term 
\cite{hui2020}. Further, usually these systems have very high 
conductivity, so even for reasonably high currents, the potential can be 
taken to be almost constant in the relevant region (which will be close 
to the sonic horizon).
 
We consider the case where the fluid is locally irrotational, so that
one can write ${\vec v} = \vec\nabla \phi$
where $\phi$ is the velocity potential which is locally well 
defined in the regions where the fluid is irrotational. We also
assume that the equation of state is barotropic so that $\rho$ is a
function of $p$ only. We can then define the specific enthalpy
\begin{equation}
h(p) = \int^p_0 \frac{dp^\prime}{\rho(p^\prime)} ~.
\end{equation}
 With this, we get $\vec\nabla h = \vec\nabla p/\rho(p)$. In terms of
$h$ and $\phi$, the Euler equation can be reduced to
\begin{equation}
\partial_t \phi + h + \frac{1}{2} (\vec\nabla \phi)^2 = 0 ~.
\end{equation}
 Consider now small perturbations $(\rho_1, p_1, \phi_1)$
on a background flow $(\rho_0, p_0, \phi_0)$. Then, it can be shown that the linearised evolution equation for $\phi_1$ can be written compactly as \cite{unruh1981, visser1998})
\begin{equation}
\partial_a (\sqrt{-g} g^{ab} \partial_b \phi_1) = 0~.
\label{eq:kg}
\end{equation}
Here, $g^{ab}$ is a matrix whose elements are functions of the background velocity, density and local speed of sound, $c_s$, in the fluid. $g_{ab}$ is the inverse matrix of $g^{ab}$ and $g=det(g_{ab})$. Notice that \cref{eq:kg} is structurally same as the relativistic wave equation for a massless scalar field ($\phi_1$) propagating in a curved spacetime with metric $g_{ab}$. Thus, we can identify $g_{ab}$ as an effective acoustic metric seen by acoustic perturbations in the velocity potential of the fluid.  It is given by the following line element:
\begin{align}
ds^2 
& \, = g_{ab} dx^a dx^b \nn \\
& \, = \Omega [-c_s^2 dt^2 +
(dx^\alpha - v^\alpha dt) (dx^\beta - v^\beta dt) \delta_{\alpha \beta}] ~.
\end{align}
$\Omega = \rho/c_s$ is a conformal factor, 
the local speed of sound, $c_s = 
\sqrt{\frac{\partial p}{\partial \rho}}$ and ${\vec v}$ is
the background flow velocity. (Here on, we use $\rho, {\vec v}, p$ etc. to denote 
 background values, without the subscript $0$. Latin alphabets $a,b,...$ denote spacetime
indices while Greek alphabets $\alpha,\beta,...$ denote spatial indices.)  
As long as the above equations of fluid dynamics and the conditions imposed on them hold good, this derivation of an acoustic metric will remain valid in any dimension. 
$\Omega, \vec v$ and $c_s$ appearing in the metric all depend on the specific nature of the electron flow, e.g. $c_s$ is determined by electron-electron interactions in the fluid.

Now, we consider an effectively one-dimensional steady flow 
of an electron fluid. We can orient the axes of our coordinate system such that the flow is parallel to the $z$ axis and the velocity vector points in the direction of decreasing $z$. So, $v^\alpha(t,z) = (0,0, - v^z(z))$  (which is irrotational). 
So, the acoustic line element simplifies to
\begin{align}
ds^2
 = 
\Omega 
\left[
- \left( c_s^2- (v^{ z})^2 \right) dt^2 
 + 2  v^{ z} dt d z            
 + dx^2 + dy^2 + d z^2
\right] ~.
\label{eq:acmet}
\end{align}
This is actually qualitatively similar to the Schwarzschild metric written in Painlev\'{e}-Gullstrand coordinates 
except that the metric coefficients here are functions of $z$ instead of the radial coordinate, as in the spherically symmetric Schwarzschild black hole. If the velocity field of the fluid is such that, given some value $z=z_H$,
\begin{align*}
v^z & \, < c_s \qquad \text{for} \ z > z_H~, \\
v^z & \, = c_s \qquad \text{for} \ z = z_H~, \\
v^z & \, > c_s \qquad \text{for} \ z < z_H~,
\end{align*}
then an acoustic horizon forms at $ z = z_H$. The fluid flowing with supersonic velocities in $z < z_H$ sweeps away all acoustic perturbations away from the horizon. The supersonic region is thus acoustically disconnected from the subsonic region. Now, if $v^z \rightarrow 0$ as $z \rightarrow z_0$ ($z_0 > z_H$), then we get back Minkowski metric there. Thus, an observer at $z=z_0$ would serve as an ``asymptotic observer'' in ``asymptotically flat'' spacetime for our purposes. (The setup described here is similar to that of \cref{fig:fig2}. For convenience, as explained later, we have adopted a different orientation of the coordinate axes in \cref{fig:dl}.) If the fluidic system were effectively two-dimensional, we would similarly get a $(2+1)$ dimensional acoustic metric with the same structure.

A remarkable property of this system is that due to the presence of a purely absorbing boundary condition at the horizon, there would be a spontaneous emission of phonons (quantised acoustic perturbations) near the horizon in the form of \textit{acoustic Hawking radiation}. This radiation is expected to be thermal and its temperature is given by,
\begin{equation}
T =  \frac{\kappa}{2\pi} = - \frac{1}{2\pi} \frac{\partial v^z}{\partial z} \bigg |_{z_H}~.     
\label{eq:ht}
\end{equation}
Here, $\kappa$ is known as the surface gravity at the acoustic horizon. After reinstating the fundamental constants $\hbar, k_B$ which had otherwise been set equal to unity, the above equation becomes,
\begin{equation}
k_B T = - \frac{\hbar}{2\pi} \frac{\partial v^z}{\partial z} \bigg |_{z_H}~.     
\label{eq:htconst}
\end{equation}
The conformal factor $\Omega$ does not affect the value of 
the temperature here \cite{das2020}. The acoustic metric has
been derived starting from fluid equations that allow the freedom to
multiply the metric by an overall constant. We can utilize this to replace the
conformal factor $\Omega(z)$ in \cref{eq:acmet}  by $\Omega(z)/\Omega (z_0)$
where $z_0$ denotes the location of the asymptotic observer.  
Since we are considering a steady 
state flow where $\Omega = \rho(z)/c_s$ is a function of $z$ only, 
the new normalised conformal factor remains unity at $z = z_0$ at all times. A conformal factor with this 
asymptotic behavior does not affect the asymptotic Hawking temperature \cite{das2020}.

In the above discussion, the starting point has been  non-relativistic
fluid equations. The relativistic fluid case would be directly relevant for the case of Dirac materials like
graphene, especially due to the expectation of very low viscosity in such 
systems. An acoustic metric can be derived for relativistic hydrodynamics
case also \cite{bilic1999,ge2010,visser2010} and the basic physics of our proposal should carry over to the relativistic regime, though we do not discuss it
in this article (see also \cite{fagnocchi2010,anacleto2010,anacleto2011,giacomelli2017,ge2019}).   Further, even for Dirac fermions in graphene,
non-relativistic fluid equations have been used as an approximation 
\cite{mayzel2019}.  So, we will work in the same spirit and assume that the basic 
physical idea behind our approach remains valid even for such systems.
 
\section{Supersonic flow of electrons in a de Laval nozzle}  \label{sec:ehydro}

\begin{figure}
\centering
\includegraphics[width=0.75\linewidth]{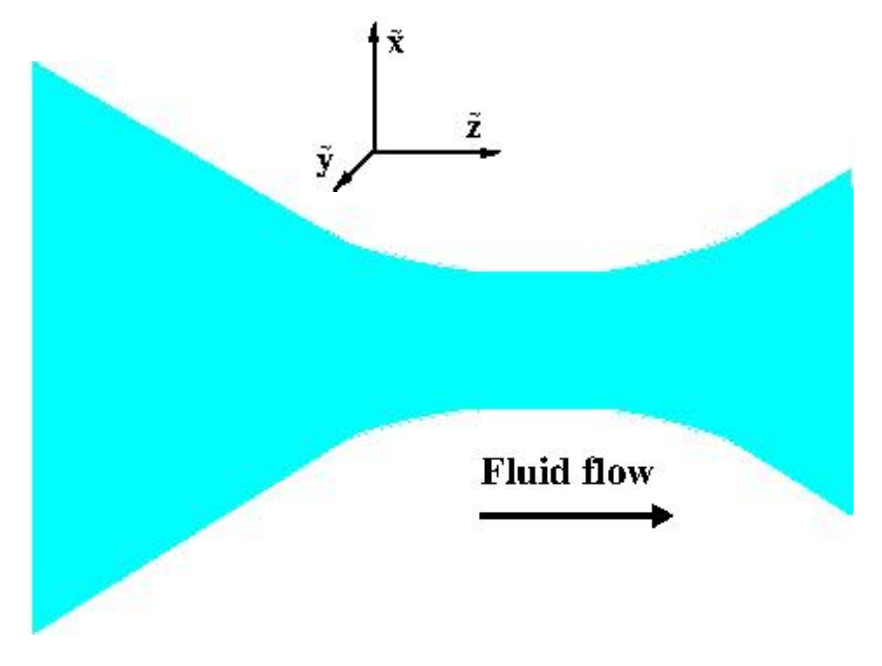}
\caption{de Laval geometry with converging-diverging flow} 
\label{fig:dl}
\end{figure}

Hydrodynamics of electrons have been extensively investigated recently and
specific experimental investigations/proposals have discussed specific
geometries of the sample, focusing on different aspects of electron
hydrodynamical flow \cite{hui2020,moors2021}. Here, we discuss the specific example of a de Laval nozzle which has a converging-diverging geometry, as shown in \cref{fig:dl} \cite{moors2021}. We  assume the flow of electron fluid  to be along the ${\tilde z}$ axis. Note that we use here  $(\tilde x, \tilde y, \tilde z)$ coordinates distinct from the $(x,y,z)$ coordinate used above in the derivation of an acoustic metric. This is because,
a discussion of acoustic black holes is most conveniently done with a
choice of z axis such that the fluid velocity vector points towards $-\hat z$
and its magnitude decreases with increasing values of
z (with fluid velocity approaching zero at some large $z = z_0$). This
is the standard convention in the literature of acoustic black holes
with 1-dimensional fluid flow. In contrast, for the discussion of fluid
flow in nozzles, eg., a de Laval one, it is traditional to choose a $\tilde z$ coordinate
such that fluid velocity increases with increasing $\tilde z$. 

The width of the quasi 2-dimensional electron gas system is taken to be 
along ${\tilde x}$ axis, and ${\tilde y}$ axis represents the thickness of
the film (which is assumed to be small). We take the ${\tilde x}$ dimension of the sample to
initially decrease along the ${\tilde z}$ axis, i.e. along the fluid flow. This will 
cause fluid velocity to increase as a function of ${\tilde z}$. 
With suitable values of system parameters, the flow can achieve sonic velocity 
 at a specific value of $\tilde z = z_H$ which represents the 
location of the horizon of the acoustic black hole. (To avoid using
too many different notations, we denote the location of the sonic horizon 
by z$_H$ throughout the article.)
It is important to  take note of the fact that within a converging or diverging shape, the flow of
electrons can not be strictly one dimensional as the flow has to converge
towards the narrowest part and diverge beyond it. Thus, flow velocity will have non-zero $v^{\tilde x}$ 
components,
which will be significant especially near the boundary of the
sample. However, near ${\tilde x} = 0$, all along the ${\tilde z}$ axis, 
$v^{\tilde x}$ 
will be negligible. Thus, our assumption of a one dimensional flow remains a reasonably
good approximation in this region and we shall neglect $v^{\tilde x}$ 
 in our discussion. (This is same as the situation encountered in
the discussion of Hawking radiation in Bose-Einstein condensates 
using a de Laval geometry in ref.\cite{barcelo2001}.)

 In the following, we derive the variation of the flow velocity along 
 ${\tilde z}$ in a de Laval geometry \cite{barcelo2001}. Let $A({\tilde z})$ be the cross-sectional area perpendicular to the ${\tilde z}$
axis. Continuity equation then gives
\be
\frac{d}{d{\tilde z}}(\rho A v) = 0 \Rightarrow \rho^\prime = - 
\rho \left[\frac{A^\prime}{A} + \frac{v^\prime}{v} \right].
\label{eq:lcont}
\ee
Here, a $\prime$ denotes derivation w.r.t. ${\tilde z}$ and
 $v = |\vec v| = v^{\tilde z}$. The fluid acceleration $a$
is given by
\be
{\vec a} \equiv \frac{d \vec v}{dt} = ({\vec v}.\vec\nabla){\vec v} ~,
\ee
since $\frac{\partial v}{\partial t} = 0$. For the 1-dimensional flow, we 
get $a = v \frac{dv}{d{\tilde z}} = v v^\prime$. With this, 
\cref{eq:lcont} becomes
\be
\rho^\prime = -\rho \left[ \frac{A^\prime}{A} + \frac{a}{v^2} \right].
\ee
Euler equation for time independent 1-dimensional flow gives
\be
\rho v \frac{dv}{d{\tilde z}} \equiv \rho a = - \frac{dp}{d{\tilde z}} = 
- \frac{dp}{d\rho} \rho^\prime~.
\ee
The last equality follows from a barotropic equation of state for the fluid. Using speed of sound $c_s^2 = \frac{dp}{d\rho}$ and eliminating 
$\rho^\prime$ from the above two equations we get
\be
a = \frac{-v^2 c_s^2}{(c_s^2 - v^2)} \frac{A^\prime}{A}~.  
\label{eq:nozzle}
\ee
This is known as the {\it Nozzle equation}. This shows that
for  a focussing geometry with $A^\prime < 0$, the fluid accelerates
as long as $v < c_s$. Further acceleration of the fluid to supersonic velocities $v>c_s$ can only be achieved if $A^\prime > 0$. This happens in the diverging part of the Laval nozzle. Truncating a de Laval nozzle at the throat where $A^\prime=0$ gives us what is know as a Venturi geometry. Though a fluid in this geometry would achieve sonic velocity at the neck giving rise to an acoustic horizon, we do not expect Hawking radiation in this configuration as it admits no supersonic region or  acoustic black hole where negative energy Hawking partner modes of phonons can be absorbed. 

Now, to evaluate fluid acceleration
$a(z_H)$ from \cref{eq:nozzle}, with $v(z_H) = c_s$ and $A^\prime(z_H) = 0$, we
use the l'Hospital rule. For simplicity, we consider the case 
of a constant speed of sound $c_s$. Then one obtains \cite{barcelo2001}
\begin{align}
a(z_H) & \, = \frac{c_s^3}{v^\prime (z_H)} \frac{A''}{2A} \bigg |_{z_H}~,
\label{eq:la} \\
\text{or,} \qquad 
\frac{dv}{d{\tilde z}} \bigg |_{z_H} & \, = c_s \sqrt{\frac{A''}{2A}\bigg |_{z_H}}~.
\label{eq:lvprime}
\end{align}
%

\section{Estimates of Hawking temperature} \label{sec:ht}
\begin{figure}
\centering
\includegraphics[width=0.9\linewidth]{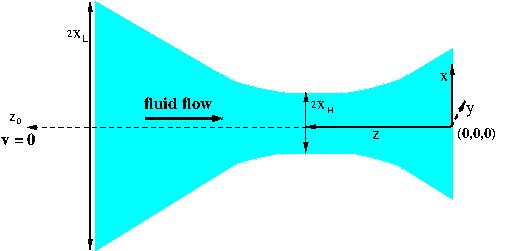}
\caption{Detailed picture of sample with de Laval geometry showing
various dimensions of the sample. The neck is assumed to have
a parabolic cross-section which should smoothly merge to the wedge shaped
geometry away from the neck. The thickness of the sample (in y-direction)
is d. The orientation of coordinate axes here are different from that of \cref{fig:dl}.}
\label{fig:fig2}
\end{figure}

We now discuss specific values of system parameters and estimate resulting 
Hawking temperature. \Cref{fig:fig2} shows a detailed picture of the proposed de Laval 
geometry where various dimensions of the sample are marked. Note that the z axis of \cref{fig:fig2} is oriented opposite to the ${\tilde z}$ axis of \cref{fig:dl}. This choice of z axis is consistent with that made in the derivation of the acoustic
black hole metric. The narrowest 
part of the neck of the nozzle, where the sonic horizon is located, is still 
denoted as $z_H$, now at $z = z_H$. Thus, \cref{eq:lvprime} written in z coordinate
becomes
\be 
\frac{dv}{dz} = - \frac{dv}{d{\tilde z}} \bigg|_{z_H} 
=  
- c_s \sqrt{\frac{A''}{2A}\bigg |_{z_H}} ~.
\label{eq:lvprime2}
\ee
With this, the Hawking temperature (restoring fundamental constants) is  given by (\cref{eq:ht})
\begin{equation}
k_B T = - \frac{\hbar}{2\pi} \frac{dv}{dz} \bigg|_{z_H} = 
- \frac{\hbar}{2\pi} c_s \sqrt{\frac{A''}{2A} \bigg |_{z_H}}~.
\label{eq:lht}
\end{equation}
(Note that $A^\prime \equiv \frac{\partial A}{\partial \tilde z} = - \frac{\partial A}{\partial z}$ and $A'' = \frac{\partial^2 A}{\partial \tilde z^2} = \frac{\partial^2 A}{\partial z^2}$.) 
The width
of the system in $x$ direction is $2 x_L$ at the left edge, and it
is $2 x_H$ at the nozzle neck, at $z = z_H$. The diagram only shows a small
part of the nozzle for the supersonic region $z < z_H$.  It seems reasonable to assume
that the supersonic region should be larger than the typical wavelength of Hawking radiation. To calculate $A''|_{z_H}$, we need to know the detailed shape of the neck.
For a simple estimate, let's assume it is of a parabolic shape at the neck which smoothly changes to a wedge shaped
geometry little away from the neck. The Hawking radiation being only
sensitive to near-horizon flow geometry, shape changes further
away from the neck do not affect the estimates of Hawking temperature. Let the
upper part of the parabolic region of this neck be characterized by,
\be
x = b (z-z_H)^2 + x_H 
\ee
Here, $b$ is a positive constant. The thickness of the quasi 2-dimensional nozzle in the $y$ direction is taken
to be $d$. Then the cross-sectional area $A_{bh}= 2 x_H d$ and   
$(A''/A)|_{z_H} = 2b/x_H$. 

For sample parameters, we take values of the same order as used in 
literature \cite{hui2020,moors2021}.  Thus, we take $x_H = 1 \mu$m and  $x_L =  
5 \mu$m.  To have a reasonable value of $a$, we consider
a parabolic shape such that when $|z-z_H| = 
2 \mu$m, we get  $x = 2 \mu$m.
This gives $b = 0.25$ ($\mu$m)$^{-1}$. For $c_s$ we take typical Fermi
velocity of electrons, $c_s \sim 10^6$m/s. Using \cref{eq:lvprime},
we get $dv/dz = 5 \times 10^{11}$s$^{-1}$. Finally, 
from \cref{eq:lht}, we
estimate the Hawking temperature to be about 0.6 K. The peak frequency
for this black body Hawking radiation is about 10$^{10}$ Hz,
corresponding to the energy of Hawking phonons.

Few points need to be discussed here. The Hawking temperature in Eqn.(26)
is the temperature that an asymptotic observer sitting in asymptotically flat spacetime would measure. The acoustic metric becomes flat when the fluid velocity
becomes zero. In our sample geometry in \cref{fig:fig2}, the left most part  
has  width $2x_L = 10 \mu$m. As the flow velocity is $v = c_s$
at the neck where the width is $2x_H = 2 \mu$m, the flow velocity
at the left edge will be $c_s/5$, directed towards the horizon. Thus there is no asymptotic observer
in our sample geometry. The observer
at the left edge of the sample sees a blueshifted Hawking temperature. Lorentz $\gamma$ factor for $v = c_s/5$
is about 1.02, thus introducing only a negligible correction to the value of
Hawking temperature.

Second point is about the effect of non-zero $v_x$ components of flow.
As we discussed above, $v_x$ will be almost zero near the $z$ axis, while
it will be significant near the sample boundaries in the $x$ direction.
Due to the fact that $A^\prime = 0$ at the sonic horizon, we expect
$v_x$ not to play a significant role in that region. However, for the
left edge, where the observer is located, one needs to restrict
attention to the region near $x = 0$ so that $v_x$ components of the
flow can be safely neglected.

\section{Observational aspects} \label{sec:obs}
The Hawking radiation here is composed of quanta of acoustic
perturbations in the velocity potential of the electron fluid. It has
a thermal spectrum with a peak frequency of about 10$^{10}$ Hz. For
observations at the left edge of the sample, we can estimate
the flux of radiation as follows.
 We first calculate the area of the horizon. For this we
need thickness $d$ of the electron gas system. 
The peak frequency of 10$^{10}$ Hz, with sound velocity $c_s = 10^6$ m/s
gives phonon wavelength of about 100 $\mu$m. For a consistent picture
of Hawking radiation, the thickness $d$ should be of order
of the peak phonon frequency, i.e. about 100 microns. However, typical
thickness of heterostructures is of the order of
few hundred nanometres, much smaller than the phonon wavelength. We shall ignore this issue for quasi 2-D materials. For 3-D materials this will not
be an issue.
In our estimates, $d$ only enters in calculating total flux of Hawking 
radiation, the area of the horizon being $A_{bh} = 2x_H d$. The power 
of Hawking radiation emitted is,
\be
P_{bh} = \sigma T^4 A_{bh} ~.
\ee
We assume that this entire power is focussed towards the left edge
of the sample, neglecting any phonon absorption at the sample
boundary. This is in the spirit of neglecting momentum transfer from
the electron fluid to the sample boundaries (necessary to get
electron hydrodynamics regime in the first place).  The flux of Hawking 
radiation obtained at the left edge is then
\be
F(z_L) = \sigma T^4 \frac{2 x_H d}{2 x_L d} = \sigma T^4 \frac{x_H}{x_L}~,
\ee
where $z_L$ denotes the $z$ coordinate at the left edge of the sample.
This flux of Hawking radiation is made up of quantized 
sound modes or phonons. Flux of
energy in a sound wave with frequency $f$ and amplitude $A_{sound}$
is given by
\be
F_{sound} = 2\pi^2 \rho c_s f^2 A_{sound}^2 ~.
\ee
$\rho = n_e m_e$ where $n_e$ and $m_e$ are electron number density
and effective electron mass in the sample. As a sample value \cite{hui2020}, 
we take $n = 10^{16}/d$ m$^{-2}$. For sample thickness, we
take  $d = 100$ nm. With this we get $n = 10^{23}$ m$^{-3}$. For 
$m_e$, we take the free electron mass. For frequency $f$, we take the
peak frequency of Hawking radiation, $f \simeq 10^{10}$ Hz. With these values, we equate the energy flux of
the sound wave to the energy flux of Hawking radiation at the left edge
of the sample to get
\be
A_{sound} \simeq 3 \times 10^{-15} m ~.
\ee
The ratio of the oscillatory part of the electric current to the 
average current at the left edge of the sample is given by
\be
\frac{I_{oscl}}{I_0} = \frac{A_{sound} f}{v(z_L)} \simeq 10^{-10} .
\ee
For a background current $I_0$ of order milliamperes, 
$I_{oscl} \sim 10^{-13}$ amperes.
It is unclear to us whether such an oscillatory current can be
observed through electromagnetic radiation. The total flux of
microwave photons in this comes out to be too small. However, one may be
able to observe this current oscillation directly. The important factor
which distinguishes this current from a general  background noise
is its black body spectrum. Further, direct dependence on parameters
like the shape of the neck and $x_H/x_L$ ratio can help in identifying
the signal.

\section{Conclusions} \label{sec:concl}
 We have proposed the possibility of observing Hawking radiation in an acoustic black hole
 system for electron hydrodynamics. This is expected to be realized in
 ultra-clean quasi 2-D materials as well as in Dirac and Weyl semi-metals
 in 3-D. For typical parameter values of such samples, our estimate
 gives a Hawking temperature of about 1K. The resulting Hawking radiation
 will manifest in terms of sound modes of the electron fluid, hence in
 electric current oscillations. We estimate amplitude of current
 oscillations to be of order $I_{oscl}/I_0 \simeq 10^{-10}$. This current
 oscillation will have strictly black body spectrum of frequency as expected
 of Hawking radiation. Its specific dependence on system parameters, such as
 the curvature of the region near sonic horizon, can help in separating
 this signal from background noise.

 We have made many strong simplifying assumptions. The peak wavelength
 of Hawking radiation has been estimated to be about 100 $\mu$m. For 
 consistency, one should require all dimensions of the sample (subsonic
 region, supersonic region, and the thickness of the sample) to have at least
 this size. For standard experimental situations, this is not the case.
 It may not be easy to prepare ultra-clean samples of this size. For smaller
 system sizes, one may expect corrections to the estimates we have provided.
 However, the qualitative picture of acoustic black hole and resulting Hawking
 radiations should remain applicable.

It will be very interesting to calculate the current-current correlations
for the subsonic and supersonic regions. This will carry signatures of
the correlations between Hawking partners, just like density-density correlations in cold
atom systems.  For the system size we
have considered, any electromagnetic radiation resulting from
current oscillations is expected to be negligible. However, with suitably chosen parameters,
e.g.  size/shape of the sample, it may be possible
to observe imprints of this Hawking radiation in electromagnetic
radiation resulting from current oscillations.

\acknowledgements{
OG would like to thank Chaitra Hegde for discussions on electron hydrodynamics and also wishes to acknowledge support from IoE-IISc fellowship.}

\bibliographystyle{eplbib}
\bibliography{ehydro}

\end{document}